\documentclass[]{svmult}

 \usepackage{graphicx}
 
\usepackage{makeidx}\makeindex
 
 \usepackage{multicol}
 \usepackage{footmisc}

 \usepackage{bm}
 \usepackage{amsmath}
 \usepackage{amssymb}
 \usepackage{dsfont}
 \usepackage{latexsym}
 \usepackage{amsfonts}
 \usepackage{epsfig}
 \usepackage{epstopdf}
 \usepackage{color}
 \definecolor{darkblue}{rgb}{0,0,.5}
 \usepackage[linktocpage, colorlinks=true, linkcolor=darkblue, citecolor=darkblue]{hyperref}
 \usepackage[all]{hypcap}

\newcommand{\C}[1]{{\cal{#1}}}
\newcommand{\bb}[1]{\textbf{#1}}
\newcommand{\h}[1]{\hat{#1}}
\newcommand{\lr}[1]{{\langle {#1}\rangle}}

\begin{document}

\title*{Controlling the stability of steady states in continuous variable quantum systems}

\author{Philipp Strasberg, Gernot Schaller, and Tobias Brandes}
\institute{Institut f\"ur Theoretische Physik, Technische Universit\"at Berlin, Hardenbergstr. 36, D-10623 Berlin, Germany}
\maketitle

\begin{abstract}
 For the paradigmatic case of the damped quantum harmonic oscillator we present two measurement-based feedback schemes 
 to control the stability of its fixed point. The first scheme feeds back a Pyragas-like time-delayed reference signal and 
 the second uses a predetermined instead of time-delayed reference signal. We show that 
 both schemes can reverse the effect of the damping by turning the stable 
 fixed point into an unstable one. Finally, by taking the classical limit $\hbar\rightarrow0$ 
 we explicitly distinguish between inherent quantum effects and effects, which would be also present in a classical 
 noisy feedback loop. In particular, we point out that the correct description of a classical particle conditioned on a 
 noisy measurement record is given by a non-linear \emph{stochastic} Fokker-Planck equation and \emph{not} a Langevin 
 equation, which has observable consequences on average as soon as feedback is considered. 
\end{abstract}

\section{Introduction}
\index{quantum feedback}\index{commutation relation}

Continuous variable quantum systems are quantum systems whose algebra is described 
by two operators $\hat x$ and $\hat p$ (usually called position and momentum), 
which obey the commutation relation $[\hat x,\hat p] = i\hbar$. 
Such systems constitute an important class of quantum systems. They do not only describe the quantum mechanical analogue 
of the motion of classical heavy particles in an external potential, but they also arise, e.g., in the quantization 
of the electromagnetic field. Understanding them is important, e.g., in quantum optics \cite{ScullyZubairyBook1997}, 
for purposes of quantum information processing \cite{BraunsteinVanLoockRMP2005, WeedbrookEtAlRMP2012}, 
or in the growing field of optomechanics \cite{AspelmeyerKippenbergMarquardtRMP2014}. 
Furthermore, due to the pioneering work of Wigner and Weyl, such systems have a well-defined classical limit 
and can be used to understand the transition from the quantum to the classical world \cite{FaddeevYakubovskiiBook2009}. 

To each quantum system there is an operator associated, called the Hamiltonian $\hat H$, which describes its energy 
and determines the dynamics of the system if it is isolated. However, in reality each system is an \emph{open} system, 
i.e., it interacts with a large environment (we call it the bath). Since the bath is so large that we cannot describe it 
in detail, it induces effects like damping, dissipation or friction, which will eventually bring the system 
to a steady state. Classically as well as quantum mechanically it is often important to be able to counteract such 
irreversible behaviour, for instance, by applying a suitably designed feedback loop. 

\begin{figure}
  \includegraphics[width=0.50\textwidth,clip=true]{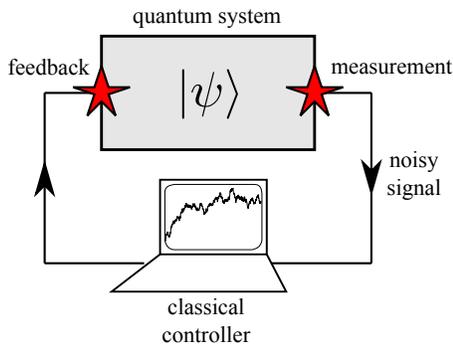}
  \label{fig quantum feedback}
  \caption{Generic sketch for a closed-loop feedback scheme, in which we wish to control the dynamics of a given quantum system. 
  Note that the feedback loop itself is actuated by a classical controller, i.e., the information after 
  the measurement is \emph{classical} (it is a number and not an operator). Nevertheless, 
  to obtain the correct dynamics of the quantum system one needs to pay additional attention 
  to the measurement and feedback step. }
 \end{figure}

In the quantum domain, however, feedback control faces additional challenges compared to the classical world 
\cite{WisemanMilburnBook2010}, see also Fig. \ref{fig quantum feedback}. 
Each closed loop control scheme starts by measuring a certain output of the system and tries to feed the 
so gained information back into the system by adjusting some system parameters to influence its dynamics. 
In the quantum world -- due to the measurement postulate of quantum mechanics and the associated ``collapse of the 
wavefunction'' -- the measurement itself significantly disturbs the system and thus, it already influences the dynamics of 
the system. If one does not take this fact correctly into account, one easily arrives at wrong conclusions. 
Nevertheless, beautiful experiments have shown that quantum feedback control is invaluable to protect quantum information 
and to stabilize non-classical states of light and matter in various settings, see e.g. Refs. 
\cite{SmithEtAlPRL2002, BushevEtAlPRL2006, GillettEtAlPRL2010, HarocheNature2011, VijayEtAlNature2012, RisteEtAlPRL2012} for 
a selection of pioneering work in this field. 

In this contribution we will apply two measurement based control schemes to a simple quantum system, the damped harmonic 
oscillator (HO), by correctly taking into account measurement and feedback noise at the quantum level (Sec. \ref{sec feedback scheme 1}
and \ref{sec feedback scheme 2}). These schemes will reverse the effect of dissipation and -- to the best of our knowledge -- 
have not been considered in this form elsewhere. However, we will see that our treatment is conceptually very close to a 
classical noisy feedback loop. With this contribution we thus also hope to provide a bridge between quantum and classical feedback 
control. For pedagogical reasons we will therefore first present the necessary technical ingredients 
(continuous quantum measurement theory, quantum feedback theory and the phase space formulation of quantum mechanics) 
in Sec. \ref{sec preliminary}. Due to a limited amount of space we cannot derive them here, but we will try to make them as 
plausible as possible. Sec. \ref{sec classical limit} is then devoted to a thorough discussion of the classical limit of our results 
showing which effects are truly quantum and which can be also expected in a classical feedback loop. 
In the last section we will give an outlook about possible applications and extensions of our feedback loop.

\section{Preliminary}
\label{sec preliminary}

\subsection{The damped quantum harmonic oscillator}
\index{harmonic oscillator, damped quantum}\index{dissipation}\index{open quantum system}\index{master equation}

We will focus only on the damped HO in this paper, but we will discuss extensions and applications 
of our scheme to other systems in Sec. \ref{sec discussion}. Using a canonical transformation we can 
rescale position and momentum such that the Hamiltonian 
of the HO reads $\hat H = \omega(\hat p^2 + \hat x^2)/2$ with $[\hat x,\hat p] = i\hbar$. 
Introducing linear combinations of position and momentum, called the annihilation operator 
$\hat a \equiv (\hat x+i\hat p)/\sqrt{2\hbar}$ and its hermitian conjugate, the creation operator $\hat a^\dagger$, 
we can express the Hamiltonian as $\hat H = \hbar\omega(\hat a^\dagger\h a + 1/2)$. Note that we explicitly keep 
Planck's contant $\hbar$ to take the classical limit ($\hbar\rightarrow0$) later on. 

The state of the HO is described by a density matrix $\h\rho$, which is a positive, hermitian operator with unit trace: 
$\mbox{tr}\h\rho = 1$. If the HO is coupled to a large bath of different oscillators at a temperature 
$T$, it is possible to derive a so-called master equation (ME) for the time evolution of the density matrix 
\cite{ScullyZubairyBook1997, WisemanMilburnBook2010, CarmichaelBook1993}: 
\begin{equation}\label{eq ME damped HO}
 \frac{\partial}{\partial t} \h\rho(t) = -\frac{i}{\hbar} [\hat H,\hat\rho(t)] + \kappa(1+n_B)\C D[\hat a]\hat\rho(t) + \kappa n_B\C D[\hat a^\dagger]\hat\rho(t).
\end{equation}
Here, we introduced the dissipator $\C D$, which is defined for an arbitrary operator $\h o$ by its action on the density matrix: 
$D[\h o]\h\rho \equiv \h o\h\rho\h o^\dagger - \{\h o^\dagger\h o,\h\rho\}/2$ where 
$\{\h a,\h b\} \equiv \h a\h b + \h b\h a$ denotes the anti-commutator. Furthermore, $\kappa > 0$ is a rate of dissipation 
characterizing how strong the time evolution of the system is effected by the bath and $n_B$ denotes the Bose-Einstein 
distribution, $n_B \equiv (e^{\beta\hbar\omega}-1)^{-1}$, where $\beta \equiv 1/T$ is the inverse temperature (we set $k_B \equiv 1$). 
For later purposes we abbreviate the whole ME (\ref{eq ME damped HO}) by 
\begin{equation}
 \frac{\partial}{\partial t} \h\rho(t) \equiv \C L_0\h\rho(t),
\end{equation}
where the ``superoperator'' $\C L_0$ is often called the Liouvillian and the subscript $0$ refers to the fact that this 
is the ME for the \emph{free} time evolution of the HO \emph{without} any measurement or feedback performed on it. 
Furthermore, it will turn out to be convenient to introduce a superoperator notation for the commutator and anti-commutator: 
\begin{equation}
 \C C[\h o]\h\rho \equiv [\h o,\h\rho], ~~~ \C A[\h o]\h\rho \equiv \{\h o,\h\rho\}.
\end{equation}

One easily verifies that the time evolution of the expectation values of position $\lr{\h x}(t) \equiv \mbox{tr}\{\h x\h\rho(t)\}$ 
and momentum $\lr{\h p}(t) \equiv \mbox{tr}\{\h p\h\rho(t)\}$ is 
\begin{equation}
 \begin{split}
  \frac{d}{dt} \lr x(t)	&=	\omega \lr p(t) - \frac{\kappa}{2} \lr x(t),	\\
  \frac{d}{dt} \lr p(t)	&=	-\omega \lr x(t) - \frac{\kappa}{2} \lr p(t),
 \end{split}
\end{equation}
as for the classical damped harmonic oscillator. More generally speaking, for an arbitrary dynamical system 
these equations describe the generic situation for a two-dimensional stable steady state $(x^*,p^*) = (0,0)$ 
within the linear approximation around $(0,0)$. For $\kappa < 0$ this would describe an unstable steady state, 
but physically we can only allow for positive $\kappa$. The positivity of $\kappa$ is mathematically required by 
Lindblad's theorem \cite{LindbladCMP1976, GoriniEtAlJMP1976} to guarantee that Eq. (\ref{eq ME damped HO}) describes 
a valid time evolution of the density matrix.\footnote{The situation of an unstable fixed point would be modeled 
by exchanging the operators $\h a$ and $\h a^\dagger$ in the dissipators. 
This would correspond to a negative $\kappa$ in the equation for the mean position and momentum. The feedback schemes 
presented here also work in that case. } 

Finally, a reader unfamiliar with this subject might find it instructive to verify that the canonical equilibrium state 
\begin{equation}
 \h \rho_\text{eq} \sim e^{-\beta\h H} \sim e^{-\beta\hbar\omega\h a^\dagger\h a} 
\end{equation}
is a steady state of the total ME (\ref{eq ME damped HO}) as it is expected from arguments of equilibrium 
statistical mechanics.

\subsection{Continuous quantum measurements}
\index{quantum measurement theory}\index{quantum measurement, weak}\index{quantum measurement, continuous}

In introductory courses on quantum mechanics (QM) one only learns about projective measurements, which yield the maximum 
information but are also maximally invasive in the sense that they project the total state $\h\rho$ onto a single 
eigenstate. QM, however, also allows for much more general measurement procedures \cite{WisemanMilburnBook2010}. 
For our purposes, so-called continuous quantum measurements are most suited. They arise by considering very 
weak (i.e., less invasive) measurements, which are repeatedly performed on the system. In the limit where the time between 
two measurements goes to zero and the measurement becomes infinitely weak, we end up with a continuous quantum 
measurement scheme. For a quick introduction see Ref. \cite{JacobsSteckContempPhys2006}. Using their notation, one needs 
to replace $k \mapsto \gamma/(4\hbar)$ to obtain our results. 

For our purposes we want to continuously measure (or ``monitor'') the position of the HO. Details of how to model 
the system-detector interaction can be found elsewhere 
\cite{BarchielliLanzProsperiNuovoCim1982, BarchielliLanzProsperiFoundPhys1983, CavesMilburnPRA1987, WisemanMilburnPRA1993, DohertyJacobsPRA1999}. 
Here, we restrict ourselves to showing the results and we try to make them plausible afterwards. 
If we neglect any contribution from $\C L_0$ for the moment, the time evolution of the density matrix due to the 
measurement of $\hat x$ is \cite{WisemanMilburnBook2010, JacobsSteckContempPhys2006, BarchielliLanzProsperiNuovoCim1982, BarchielliLanzProsperiFoundPhys1983, CavesMilburnPRA1987, WisemanMilburnPRA1993, DohertyJacobsPRA1999}
\begin{equation}\label{eq ME av meas}
 \frac{\partial}{\partial t} \h\rho(t) = -\frac{\gamma}{4\hbar}\C C^2[\h x]\h\rho(t) \equiv \C L_\text{meas}\h\rho(t).
\end{equation}
Here, the new parameter $\gamma$ has the physical dimension of a rate and quantifies the strength of the measurement. 
For $\gamma = 0$ we thus recover the case without any measurement. It is instructive to have a look how the 
matrix elements of $\h\rho(t)$ evolve in the measurement basis $|x\rangle$ of the position operator 
$\h x = \int dx x|x\rangle\langle x|$: 
\begin{equation}
 \frac{\partial}{\partial t}\langle x|\h\rho(t)|x'\rangle = - \frac{\gamma}{4\hbar}(x-x')^2 \langle x|\h\rho(t)|x'\rangle.
\end{equation}
We thus see that the off-diagonal elements (or ``coherences'') are exponentially damped whereas the diagonal elements 
(or ``populations'') remain unaffected. This is exactly what we would expect from a weak quantum measurement: the density 
matrix is perturbed only slightly but finally, in the long-time limit, it becomes diagonal in the measurement basis. 
Note that in case of a standard projective measurement scheme, the coherences would \emph{instantaneously vanish}. 

\index{density matrix, conditional}

The ME (\ref{eq ME av meas}) is, however, only half of the story, because it tells us only about the average time evolution 
of the system, i.e., about the whole ensemble $\h\rho$ averaged over all possible measurement records. The distinguishing 
feature of closed-loop control (as compared to open-loop control) is, however, that we want to influence the system 
based on a \emph{single} (and not ensemble) measurement record. We denote the density matrix conditioned on a certain measurement 
record by $\h\rho_c$ and call it the conditional density matrix. 
Its classical counterpart would be simply a conditional probability distribution. 

\index{Wiener increment}

In QM, even in absence of classical measurement errors, each single measurement record is necessarily noisy 
due to the inherent probabilistic interpretation of measurement outcomes in QM. 
The measurement signal $I(t)$ associated to the continuous position measurement scheme above can be shown to obey the stochastic process 
\cite{WisemanMilburnBook2010, JacobsSteckContempPhys2006} 
\begin{equation}\label{eq meas record}
 dI(t) = \langle\h x\rangle_c(t)dt + \sqrt{\frac{\hbar}{2\gamma\eta}} dW(t).
\end{equation}
Here, by $\langle\h x\rangle_c(t)$ we denoted the expectation value with respect to the conditional density matrix, i.e., 
$\langle\h x\rangle_c(t) \equiv \mbox{tr}\{\h x\h\rho_c(t)\}$. Furthermore, $dW(t)$ is the Wiener increment. 
According to the standard rules of stochastic calculus, it obeys the relations 
\cite{WisemanMilburnBook2010, JacobsSteckContempPhys2006} 
\begin{equation}
 \mathbb{E}[dW(t)] = 0, ~~~ dW(t)^2 = dt
\end{equation}
where $\mathbb{E}[\dots]$ denotes a (classical) ensemble average over all noisy realizations. 
Furthermore, we have introduced a new parameter $\eta\in[0,1]$, which is used to model the efficiency of the detector 
\cite{WisemanMilburnBook2010, JacobsSteckContempPhys2006, WisemanMilburnPRA1993} with $\eta = 1$ corresponding to the 
case of a perfect detector. 

\index{master equation, stochastic}

Finally, we need to know how the state of the system evolves conditioned on a certain measurement record. 
This evolution is necessarily stochastic due to the stochastic measurement record. The so-called stochastic ME (SME) turns out to 
be given by \cite{WisemanMilburnBook2010, JacobsSteckContempPhys2006} 
\begin{equation}\label{eq SME meas}
 \h\rho_c(t+dt) = \h\rho_c(t) + \C L_\text{meas}\h\rho_c(t)dt + \sqrt{\frac{\gamma\eta}{2\hbar}} \C A[\h x-\langle\h x\rangle_c(t)]\h\rho_c(t) dW(t).
\end{equation}
Because it will turn out to be useful, we have written the SME in an ``incremental form'' by explicitly using differentials as one 
would also do for numerical simulations. By definition we regard the quantity 
$[\h\rho_c(t+dt)-\h\rho_c(t)]/dt$ as being equivalent to $\partial_t \h\rho_c(t)$. 
Using Eq. (\ref{eq meas record}) we can express the SME alternatively as 
\begin{equation}
 \h\rho_c(t+dt) = \h\rho_c(t) + \C L_\text{meas}\h\rho_c(t)dt + \frac{\gamma\eta}{\hbar} \C A[\h x-\langle\h x\rangle_c(t)]\h\rho_c(t) [dI(t) - \langle\h x\rangle_c(t)dt],
\end{equation}
which explicitly demonstrates how our knowledge about the state of the system changes conditioned on a given measurement 
record $I(t)$.\footnote{We explicitly adopt a Bayesian probability theory point of view in which probabilities (or more 
generally the density matrix $\h\rho$) describe only (missing) human information. Especially, different observers (with 
possibly different access to measurement records) would associate \emph{different} states $\h\rho$ to the \emph{same} 
system. } 
We remark that the SME for $\h\rho_c(t)$ is nonlinear in $\h\rho_c(t)$, due to the fact that this in an 
equation of motion for a \emph{conditional} density matrix. 

\index{It\^o calculus}

To obtain the ME (\ref{eq ME av meas}) for the average evolution, we only need to average the SME (\ref{eq SME meas}) 
over all possible measurement trajectories. In fact, it can be shown (see \cite{WisemanMilburnBook2010, JacobsSteckContempPhys2006}) 
that Eq. (\ref{eq SME meas}) has to be interpreted within the rules of It\^o stochastic calculus (as well as all the following 
stochastic equations unless otherwise mentioned) such that 
\begin{equation}\label{eq Ito average}
 \mathbb{E}[\h\rho_c(t)dW(t)] = \mathbb{E}[\h\rho_c(t)]\mathbb{E}[dW(t)] = 0
\end{equation}
holds. Defining $\h\rho(t) \equiv \mathbb{E}[\h\rho_c(t)]$, one can readily verify that the SME (\ref{eq SME meas}) yields 
on average Eq. (\ref{eq ME av meas}). 

\index{quantum trajectory}

Taking the free evolution of the HO into account, Eq. (\ref{eq ME damped HO}), the total stochastic evolution of the system obeys 
\begin{equation}\label{eq SME free plus meas}
 \h\rho_c(t+dt) = \left\{1 + (\C L_0 + \C L_\text{meas})dt + \sqrt{\frac{\gamma\eta}{2\hbar}} \C A[\h x-\langle\h x\rangle_c(t)] dW(t)\right\}\h\rho_c(t).
\end{equation}
Note that there are no ``mixed terms'' from the free evolution and the evolution due to the measurement to lowest order in $dt$. 
Furthermore, we remark that a solution of a SME is called a \emph{quantum trajectory} in the literature 
\cite{WisemanMilburnBook2010, CarmichaelBook1993, GardinerZollerBook2004}.

\subsection{Direct quantum feedback}
\index{quantum feedback, direct}

In the following we will consider a form of quantum feedback control, which is sometimes called direct quantum feedback 
control, because the measurement signal is directly fed back into the system (possibly with a delay) without any additional 
post-processing of the signal as, e.g., filtering or parameter estimation \cite{DohertyJacobsPRA1999}. 
Direct quantum feedback based on a continuous measurement scheme was developed by Wiseman and Milburn 
\cite{WisemanMilburnPRL1993, WisemanMilburnPRA1994, WisemanPRA1994}. Experimentally, the idea would be to continuously 
adjust a parameter of the Hamiltonian based on the measurement outcome (\ref{eq meas record}) to control the dynamics of the 
system. Theoretically, we define the feedback control superoperator $\C F$ 
\begin{equation}\label{eq feedback superoperator}
 [\dot\rho_c(t)]_\text{fb} = \C F\h\rho_c(t) \equiv -\frac{i}{\hbar}\frac{dI(t)}{dt}\C C[\h z]\h\rho_c(t),
\end{equation}
which describes a change of the free system Hamiltonian $\h H = \omega(\h p^2 + \h x^2)/2$ to a new effective Hamiltonian 
$\omega(\h p^2 + \h x^2)/2 + \frac{dI(t)}{dt}\h z$ containing a new term proportional to the measurement result and an arbitrary 
hermitian operator $\h z$ (with units of $\dot x$). Here, we neglected any delay and assumed an instantaneous feedback of the 
measurement signal, but a delay can be easily incorporated, too, see Sec. \ref{sec feedback scheme 1}. 

Because the action of the feedback superoperator $\C F$ on the system was merely postulated, we do not a priori know whether we have to 
interpret it according to the It\^o or Stratonovich rules of stochastic calculus, but it turns out that only the latter interpretation gives 
senseful results \cite{WisemanMilburnBook2010, WisemanMilburnPRL1993, WisemanMilburnPRA1994}. 
Then, the effect of the feedback on the total time evolution of the system (including the measurement and free time evolution) can 
be found by exponentiating Eq. (\ref{eq feedback superoperator}) \cite{WisemanMilburnBook2010, WisemanMilburnPRL1993, WisemanMilburnPRA1994} 
\begin{equation}\label{eq SME influence fb}
 \h\rho_c(t+dt) = e^{\C F dt}\left\{1 + \C L_0 dt + \C L_\text{meas}dt + \sqrt{\frac{\gamma\eta}{2\hbar}} \C H[\h x] dW(t)\right\}\h\rho_c(t)
\end{equation}
and this equation is again of It\^o type. 
Note that by construction this equation assures that the feedback step happens \emph{after} the measurement as it must due to causality. 
Now, expanding $e^{\C F dt}$ to first order in $dt$ with $dI(t)$ from Eq. (\ref{eq meas record}) 
(note that this requires to expand the exponential function up to \emph{second} order 
due to the contribution from $dW(t)^2 = dt$) and using the rules of stochastic calculus gives the effective SME under feedback control: 
\begin{equation}
 \begin{split}
  \h\rho_c(t+dt)	=&~	\h\rho_c(t) + dt\left\{\C L_0 + \C L_\text{meas} - \frac{i}{2\hbar}\C C[\h z]\C A[\h x] - \frac{1}{4\hbar\gamma\eta}\C C^2[\h z]\right\}\h\rho_c(t)	\\
			&+	dW(t)\left\{\sqrt{\frac{\gamma\eta}{2\hbar}} \C A[\h x-\langle\h x\rangle_c(t)] - \frac{i}{\hbar} \sqrt{\frac{\hbar}{2\gamma\eta}} \C C[\h z]\right\}\h\rho_c(t).
 \end{split}
\end{equation}
If we take the ensemble average over the measurement records, we obtain the effective feedback ME 
\begin{equation}\label{eq ME free meas fb}
 \frac{\partial}{\partial t} \h\rho(t) = \left\{\C L_0 + \C L_\text{meas} - \frac{i}{2\hbar}\C C[\h z]\C A[\h x] - \frac{1}{4\hbar\gamma\eta}\C C^2[\h z]\right\}\h\rho(t)
\end{equation}
or more explicitly for our model 
\begin{equation}
 \begin{split}
  \frac{\partial}{\partial t} \h\rho(t)	=&	-\frac{i}{\hbar}\left\{[\h H,\h\rho(t)] + \frac{1}{2}[\h z,\h x\h\rho(t) + \h\rho(t)\h x]\right\}	\\
					&+	\kappa(1+n_B)\C D[\hat a]\hat\rho(t) + \kappa n_B\C D[\hat a^\dagger]\hat\rho(t)	\\
					&-	\frac{\gamma}{4\hbar}[\h x,[\h x,\h\rho(t)]] - \frac{1}{4\hbar\gamma\eta}[\h z,[\h z,\h\rho(t)]].
 \end{split}
\end{equation}
Note that this equation is again linear in $\h\rho(t)$ as it must be for a consistent statistical interpretation. 

Before we give a short review about the last technically ingredient we need, which is rather unrelated to the previous content, 
we give a short summary. We have introduced the ME (\ref{eq ME damped HO})
for a HO of frequency $\omega$, which is damped at a rate $\kappa$ 
due to the interaction with a heat bath at inverse temperature $\beta$. We then started to continuously monitor the system at a 
rate $\gamma$ with a detector of efficiency $\eta$. This procedure gave rise to a SME (\ref{eq SME free plus meas}) 
conditioned on the measurement record (\ref{eq meas record}). Finally, we applied feedback control by instantaneously 
changing the system Hamiltonian using the operator $\h z$, which resulted in the effective ME (\ref{eq ME free meas fb}).

\subsection{Quantum mechanics in phase space}
\label{sec qm phase space}
\index{quantum mechanics in phase space}\index{Wigner function}\index{Fokker-Planck equation}\index{classical limit}

The phase space formulation of QM is an equivalent formulation of QM, in which one tries 
to treat position and momentum on an equal footing (in contrast, in the Schr\"odinger formulation one has to work either in 
the position or (``exclusive or'') momentum representation). By its design, phase space QM is very close to the classical 
phase space formulation of Hamiltonian mechanics and it is a versatile tool for a number of problems. For a more thorough 
introduction the reader is referred to Refs. 
\cite{ScullyZubairyBook1997, FaddeevYakubovskiiBook2009, GardinerZollerBook2004, HillaryEtAlPhysRep1984, CarmichaelBook1999a, ZachosFairlieCurtrightBook2005}. 

The central concept is to map the density matrix $\h\rho$ to an object called the Wigner function: 
\begin{equation}\label{eq def Wigner fct}
 W(x,p) \equiv \frac{1}{\hbar\pi} \int_{-\infty}^\infty dy \langle x-y|\h\rho|x+y\rangle e^{2ipy/\hbar}.
\end{equation}
The Wigner function is a quasi-probability distribution meaning that it is properly normalized, 
$\int dx dp W(x,p) = 1$, but can take on negative values. The expectation value of any function 
$F(x,p)$ in phase space can be computed via 
\begin{equation}
 \langle F(x,p)\rangle = \int\limits_{\mathbb{R}^2} dx dp F(x,p) W(x,p) = \mbox{tr}\{\h f(\h x,\h p) \h\rho\},
\end{equation}
where the associated operator-valued observable $\h f(\h x,\h p)$ can be obtained from $F(x,p)$ via the Wigner-Weyl transform 
\cite{FaddeevYakubovskiiBook2009, GardinerZollerBook2004, ZachosFairlieCurtrightBook2005}. 
Roughly speaking this transformation symmetrizes all operator valued expressions. 
For instance, if $F(x,p) = xp$, then $\h f(\h x,\h p) = (\h x\h p + \h p\h x)/2$. 

Each ME for a continuous variable quantum system can now be transformed to a corresponding equation of motion 
for the Wigner function. This is done by using certain correspondence rules between operator valued expressions 
and their phase space counterpart, e.g., 
\begin{equation}
 \h x\h \rho \leftrightarrow \left(x+\frac{i\hbar}{2}\frac{\partial}{\partial p}\right) W(x,p),
\end{equation}
which can be verified by applying Eq. (\ref{eq def Wigner fct}) to $\h x\h\rho$ and some algebraic manipulations 
\cite{GardinerZollerBook2004, CarmichaelBook1999a}. 

The big advantage of the phase space formulation of QM is now that many MEs (namely those which can be called ``linear'') 
transform into an ordinary Fokker-Planck equation (FPE), for which many solution techniques are known \cite{RiskenBook1984}. 
We denote the general FPE for two variables $(x,p)$ as 
\begin{equation}\label{eq FPE generic}
 \frac{\partial}{\partial t} W(x,p,t) = \left\{-\nabla^T\cdot\textbf{d} + \frac{1}{2}\nabla^T\cdot D\cdot\nabla\right\}W(x,p,t)
\end{equation}
where $\nabla^T \equiv (\partial_x, \partial_p)$, the dot denotes a matrix product, $\textbf{d}$ is the drift vector and 
$D$ the diffusion matrix. It is then straightforward to confirm that the ME (\ref{eq ME damped HO}) corresponds to 
a FPE with
\begin{equation}
 \begin{split}\label{eq W drift diffusion free HO}
  \textbf{d}_x	&=	\omega p - \frac{\kappa}{2} x, ~~~ \textbf{d}_p = -\omega x - \frac{\kappa}{2} p,	\\
  D_{xx}		&=	D_{pp} = \kappa\hbar\frac{1+2n_B}{2}, ~~~ D_{xp} = D_{px} = 0.
 \end{split}
\end{equation}
The SME (\ref{eq SME meas}) instead transforms to an equation for the conditional Wigner function $W_c(x,p,t)$ and reads 
\begin{equation}\label{eq W meas}
 W_c(x,p,t+dt) = \left\{1 + dt\frac{\gamma\hbar}{4}\frac{\partial^2}{\partial p^2} + dW(t)\sqrt{\frac{2\gamma\eta}{\hbar}}[x-\langle x\rangle_c]\right\} W_c(x,p,t).
\end{equation}
This does not have the standard form of a FPE. The additional term, however, does not cause any trouble in the interpretation 
of the Wigner function because we can still confirm that $\int dx dp W_c(x,p,t) = 1$. 

Finally, we point out that the transition from quantum to classical physics is mathematically accomplished by the limit 
$\hbar\rightarrow0$ \cite{FaddeevYakubovskiiBook2009}. 
Physically, of course, we do not have $\hbar = 0$ but the classical action of the particles motion 
becomes large compared to $\hbar$. We will discuss the classical limit of our equations in detail in Sec. \ref{sec classical limit}.

\section{Feedback Scheme I}
\label{sec feedback scheme 1}
\index{Pyragas control, quantum}

The first feedback scheme we consider is the quantum analogue of the classical scheme considered in Ref. \cite{HoevelSchoellPRE2005}. 
There the authors used a time delayed reference signal of the form (without any noise) 
\begin{equation}\label{eq meas signal pyragas}
 \begin{split}
  \delta I(t,\tau)	&\equiv	[I(t) - I(t-\tau)]dt	\\
			&=	[\langle\h x\rangle(t)-\langle\h x\rangle_\tau(t)]dt + \sqrt{\frac{\hbar}{2\gamma\eta}} [dW(t)-dW_\tau(t)]
 \end{split}
\end{equation}
to control the stability of the fixed point.\footnote{In fact, in Ref. \cite{HoevelSchoellPRE2005} they did not only feed back the results 
from a position measurement, but also from a momentum measurement. The simultaneous weak measurement of position and momentum can be also 
incorporated into our framework \cite{BarchielliLanzProsperiNuovoCim1982, ArthursKellyBSTJ1965, ScottMilburnPRA2001}, 
but this would merely add additional terms without changing the overall message.} 
Here, a subscript $\tau$ indicates a shift of the time argument, i.e., 
$f_\tau(t) \equiv f(t-\tau)$. Due to this special form such feedback schemes are sometimes called Pyragas-like 
feedback schemes \cite{PyragasPLA1992}. It should be noted however that we do not have a chaotic system here and we do not want 
to stabilize an unstable periodic orbit. In this respect, our feedback scheme is still an \emph{invasive} feedback scheme, 
because the feedback-generated force does not vanish even if our goal to reverse the effect of the damping was achieved. 
We emphasize that such feedback schemes are widely used in classical control theory to influence the behaviour of, e.g., 
chaotic systems or complex networks \cite{SchoellSchusterBook2007, FlunkertFischerSchoellBook2013} 
and quite recently, there has been also a considerable interest to explore its quantum implications 
\cite{WhalenEtAlConfProc2011, CarmeleEtAlPRL2013, SchulzeEtAlPRA2014, HeinEtAlPRL2014, GrimsmoEtAlNJP2014, KopylovEtAlNJP2015, GrimsmoArXiv2015, KabussArXiv2015}. 
However, except of the feedback scheme in Ref. \cite{KopylovEtAlNJP2015}, the feedback schemes above were designed as 
\emph{all-optical} or \emph{coherent} control schemes, in which the system is not subjected to an explicit measurement, 
but the environment is suitable engineered such that it acts back on the system in a very specific way. We will 
compare our scheme (which is based on explicit measurements) with these schemes towards the end of this section. 

To see how our feedback scheme influences our system, we can still use Eq. (\ref{eq SME influence fb}) 
together with the measurement signal (\ref{eq meas signal pyragas}). Choosing $\h z = k\h p$ with $k\in\mathbb{R}$ 
and using that $dW(t)dW_\tau(t) = 0$ for $\tau\neq 0$ we obtain the SME 
\begin{equation}
 \begin{split}\label{eq SME fb scheme 1}
  & \h\rho_c(t+dt) = \left\{1 + dt[\C L_0 + \C L_\text{meas}]\right\}\h\rho_c(t)	\\
  & - dt\left\{\frac{ik}{2\hbar}\C C[\h p]\C A[\h x-\langle x\rangle_c(t)] - \frac{ik}{\hbar}[\langle\h x\rangle(t) - \langle\h x\rangle_\tau(t)]\C C[\h p] - \frac{k^2}{2\hbar\gamma\eta}\C C^2[\h p]\right\}\h\rho_c(t)	\\
  & + dW(t) \sqrt{\frac{\gamma\eta}{2\hbar}}\C A[\h x-\langle x\rangle_c(t)]\h\rho_c(t) - \frac{ik}{\sqrt{2\hbar\gamma\eta}} [dW(t)-dW_\tau(t)] \C C[\h p] \h\rho_c(t).
 \end{split}
\end{equation}
It is important to emphasize that also \emph{time-delayed noise} enters the equation of motion for $\h\rho_c(t)$. \index{noise, time-delayed}
Because we do not know what $\mathbb{E}[\h\rho_c(t) dW_\tau(t)]$ is in general, there is \emph{a priori} no ME for the average 
time evolution of $\h\rho(t)$. Approximating $\mathbb{E}[\h\rho_c(t) dW_\tau(t)] \approx 0$ yields nonsense (the resulting ME would 
not even be linear in $\h\rho$). 
This is, however, not a quantum feature and is equally true for classical feedback control 
based on a noisy, time-delayed measurement record (also see Sec. \ref{sec classical limit}). 

Due to the fact that there is no average ME, we are in principle doomed to simulated the SME (\ref{eq SME fb scheme 1}) 
and average afterwards. However, as it turns out Eq. (\ref{eq SME fb scheme 1}) can be transformed into a stochastic FPE 
whose solution is expected to be a Gaussian probability distribution. We will then see that the covariances indeed evolve 
\emph{deterministicly}. Furthermore, it is possible to analytically deduce the equation of motion for the mean values 
on average. Within the Gaussian approximation we then have full knowledge about the evolution of the system. 

Using the results from Sec. \ref{sec qm phase space} we obtain 
\begin{equation}
 \begin{split}
  & W_c(x,p,t+dt) = W_c(x,p,t) + dt\left(-\nabla^T\cdot\bb{d} + \frac{1}{2}\nabla^T\cdot D\cdot\nabla\right) W_c(x,p,t)	\\
  & + \left\{\sqrt{\frac{2\gamma\eta}{\hbar}} dW (x-\langle x\rangle_c) - k\sqrt{\frac{\hbar}{2\gamma\eta}}(dW-dW_\tau)\frac{\partial}{\partial x}\right\} W_c(x,p,t)
 \end{split}
\end{equation}
with the nonvanishing coefficients 
\begin{align}
 \textbf{d}_x	&=	\omega p - \frac{\kappa}{2} x + k[x-\langle x\rangle_{c,\tau}(t)], ~~~ \textbf{d}_p = -\omega x - \frac{\kappa}{2} p,	\\
 D_{xx}		&=	\kappa\hbar\frac{1+2n_B}{2} + \frac{\hbar k^2}{\gamma\eta}, ~~~ D_{pp} = \kappa\hbar\frac{1+2n_B}{2} + \frac{\hbar\gamma}{2}.
\end{align}

\begin{figure}
 \includegraphics[width=0.85\textwidth,clip=true]{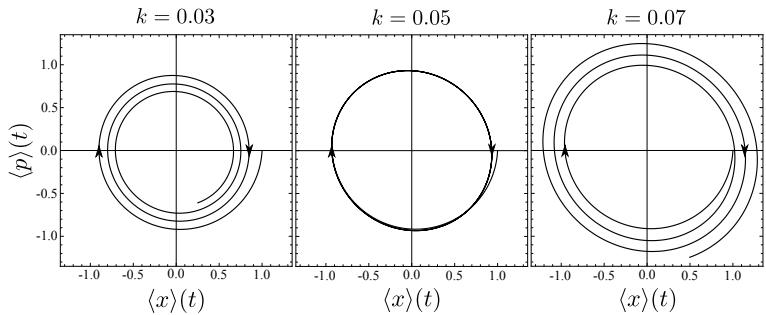}
 \label{fig plot scheme 1}
 \caption{Parametric plot of $(\lr x,\lr p)(t)$ as a function of time $t\in[0,20]$ for different feedback strengths $k$ 
 based on Eqs. (\ref{eq mean x scheme 1}) and (\ref{eq mean p scheme 1}). 
 The initial condition is $(\lr x,\lr p)(t) = (1,0)$ for $t\le 0$ and the other parameters are $\omega = 1, \kappa = 0.1$ 
 and $\tau = \pi$. Note that the trajectory for $k=\kappa$ is not a perfect circle due to the asymmetric feedback, which 
 is only applied to the $x$-coordinate and not to $p$. }
\end{figure}

We introduce the conditional covariances by
\begin{equation}
 V_{x,c} \equiv \langle x^2\rangle_c - \langle x\rangle_c^2, ~~~ V_{p,c} \equiv \langle p^2\rangle_c - \langle p\rangle_c^2, ~~~ C_{c} \equiv \langle xp\rangle_c - \langle x\rangle_c\langle p\rangle_c
\end{equation}
where we dropped already any time argument for 
notational convenience (we keep the subscript $\tau$ to denote the time-delay though). 
The time-evolution of the conditional means is then given by 
\begin{align}
 d\langle x\rangle_c	=&~	\left\{\omega\lr p_c - \frac{\kappa}{2}\lr x_c + k(\lr x_c - \lr x_{c,\tau})\right\}dt	\label{eq cond mean x scheme 1}	\\
			&+	k\sqrt{\frac{\hbar}{2\gamma\eta}} (dW-dW_\tau) + \sqrt{\frac{2\gamma\eta}{\hbar}} V_{x,c} dW,	\nonumber	\\
 d\langle p\rangle_c	=&~	\left\{-\omega\lr x_c - \frac{\kappa}{2}\lr p_c\right\}dt + \sqrt{\frac{2\gamma\eta}{\hbar}} C_{c} dW	\label{eq cond mean p scheme 1}
\end{align}
Note that -- for a stochastic simulation of these equations -- we are required to simulate the equations for the covariances 
[Eqs. (\ref{eq Vxc scheme 1}) -- (\ref{eq Cc scheme 1})], too. Interestingly however, because the time-delayed noise enters only additively, 
we can also average these equations to obtain the unconditional evolution of the mean values directly: 
\begin{align}
 \frac{d}{dt}\langle x\rangle	=&~	\omega\lr p - \frac{\kappa}{2}\lr x + k(\lr x - \lr x_{\tau}),	\label{eq mean x scheme 1}	\\
 \frac{d}{dt}\langle p\rangle	=&~	-\omega\lr x - \frac{\kappa}{2}\lr p.	\label{eq mean p scheme 1}
\end{align}
These equations are exactly the same as the classical equations in Ref. \cite{HoevelSchoellPRE2005} if one considers only 
position measurements. Hence, we can successfully reproduce the classical feedback scheme \emph{on average}. 
Unfortunately the treatment of delay differential equations is very complicated and our goal is not to study these 
equations in detail now. However, the reasoning why we can turn a stable fixed point into an unstable one goes like this: 
for $k=0$ we clearly have a stable fixed point but for $k\gg\kappa$ we might neglect the term $-\frac{\kappa}{2}\lr x$ for 
a moment. If we choose $\tau = \pi/\omega$ (corresponding to half of a period of the undamped HO), we see that the 
``feedback force'' $k(\lr x - \lr x_{\tau})$ is always positive if $\lr x > 0$ and negative if $\lr x < 0$ (we assume 
$k>0$). Hence, by looking at the differential equation it follows that the feedback term generates a 
drift ``outwards'', i.e., away from the fixed point $(0,0)$, which at some point also cannot be compensated anymore by the friction of 
the momentum $-\frac{\kappa}{2}\lr p$. From the numerics, see Fig. \ref{fig plot scheme 1}, we infer that the critical feedback strength, 
which turns the stable fixed point into an unstable one is $k\ge\frac{\kappa}{2}$, also see Ref. \cite{HoevelSchoellPRE2005} for a 
more detailed discussion of the domain of control. 

\index{Gaussian state, quantum}

Turning to the time evolution of the conditional covariances we obtain \footnote{Pay attention to the fact that 
we are using an It\^o stochastic differential equation where the ordinary chain rule of differentiation does not apply. 
Instead, we have for instance for the stochastic change of the position variance 
$dV_{x,c} = d\langle x^2\rangle_c - 2\lr x_c d\lr x_c - (d\lr x_c)^2$. }
\begin{align}
 d V_{x,c}	=&~	\left\{- \kappa V_{x,c} + 2\omega C_c - \frac{2\gamma\eta}{\hbar}V_{x,c}^2 + \kappa\hbar\frac{1+2n_B}{2}\right\} dt	\nonumber	\\
		&+	\sqrt{\frac{2\gamma\eta}{\hbar}} \left\langle (x-\lr x_c)^3\right\rangle_c dW,	\label{eq Vxc scheme 1}	\\
 d V_{p,c}	=&	\left\{-\kappa V_{p,c} - 2\omega C_c - \frac{2\gamma\eta}{\hbar}C_c^2 + \kappa\hbar\frac{1+2n_B}{2} + \frac{\hbar\gamma}{2}\right\} dt	\nonumber	\\
		&+	\sqrt{\frac{2\gamma\eta}{\hbar}} \left\langle(x-\lr x_c)(p-\lr p_c)^2\right\rangle_c dW,	\\
 d C_{c}	=&~	\left\{\omega(V_{p,c}-V_{x,c}) - \kappa C_c - \frac{2\gamma\eta}{\hbar} V_{x,c} C_c\right\} dt	\nonumber	\\
		&+	\sqrt{\frac{2\gamma\eta}{\hbar}} \left\langle(x-\lr x_c)^2(p-\lr p_c)\right\rangle_c dW.	\label{eq Cc scheme 1}
\end{align}
Unfortunately, we see that all the stochastic terms proportional to $dW$ 
involve third order cumulants, which would in turn require to 
deduce equations for them as well. However, if we assume that the state of our system is Gaussian, these terms 
vanish due to the fact that third order cumulants of a Gaussian are zero. In fact, the assumption 
of a Gaussian state seems reasonable\footnote{We remark that a Gaussian 
state in QM, i.e., a system described by a Gaussian Wigner function, might still exhibit true quantum features like 
entanglement or squeezing \cite{BraunsteinVanLoockRMP2005, WeedbrookEtAlRMP2012}. }: 
first of all, if the system is already Gaussian, it will also remain Gaussian for all times, because then the Eqs. 
(\ref{eq cond mean x scheme 1}) and (\ref{eq cond mean p scheme 1}) as well as Eqs. (\ref{eq Vxc scheme 1}) -- 
(\ref{eq Cc scheme 1}) form a closed set. Second, even if we start with a non-Gaussian distribution, 
the state is expected to rapidly evolve to a Gaussian due to the continuous position measurement and the environmentally 
induced decoherence and dissipation \cite{ZurekHabibPazPRL1993}. Then, the time evolution of the conditional covariances 
becomes indeed deterministic, i.e., the covariances (but not the means) behave identically in each single realization of 
the experiment: 
\begin{align}
 \frac{d}{dt} V_{x,c}	=&	- \kappa V_{x,c} + 2\omega C_c - \frac{2\gamma\eta}{\hbar}V_{x,c}^2 + \kappa\hbar\frac{1+2n_B}{2},	\\
 \frac{d}{dt} V_{p,c}	=&	-\kappa V_{p,c} - 2\omega C_c - \frac{2\gamma\eta}{\hbar}C_c^2 + \kappa\hbar\frac{1+2n_B}{2} + \frac{\hbar\gamma}{2},	\\
 \frac{d}{dt} C_{c}	=&	\omega(V_{p,c}-V_{x,c}) - \kappa C_c - \frac{2\gamma\eta}{\hbar} V_{x,c} C_c.
\end{align}
Thus, we can fully solve the conditional dynamics of the system by first solving the ordinary differential equations for 
the covariances and then, using this solution, we can integrate the stochastic equations (\ref{eq cond mean x scheme 1}) and 
(\ref{eq cond mean p scheme 1}) for the means. 

Because the time evolution of the conditional covariances is the same for the second feedback scheme, we will discuss 
them in more detail in Sec. \ref{sec feedback scheme 2}. 
Here, we just want to emphasize that we cannot simply average the conditional covariances to obtain the unconditional 
ones, i.e., $\mathbb{E}[V_{x,c}] \neq V_x \equiv \int dx dp x^2 W(x,p) - [\int dx dp x W(x,p)]^2$ in general. 
In fact, the conditional and unconditional covariances can behave very differently, see Sec. \ref{sec feedback scheme 2}. 

\index{quantum feedback, coherent}

Finally, let us say a few words about our feedback scheme in comparison with the coherent control schemes in Refs. 
\cite{WhalenEtAlConfProc2011, CarmeleEtAlPRL2013, SchulzeEtAlPRA2014, HeinEtAlPRL2014, GrimsmoEtAlNJP2014, GrimsmoArXiv2015, KabussArXiv2015}, 
which are designed for quantum optical systems and use an external mirror to induce an intrinsic time-delay in the system dynamics. 
Clearly, the advantage of the coherent control schemes is that they do not introduce additional noise, because they avoid any explicit 
measurement. On the other hand, in our feedback loop we have the freedom to choose the feedback strength $k$ 
at our will, which allows us to truely reverse the effect of dissipation. In fact, due to simple arguments of energy 
conservation, the coherent control schemes can only fully reverse the effect of dissipation if the external mirrors are 
\emph{perfect}. Otherwise the overall system and controller is still loosing energy at a finite rate such that the system 
ends up in the same steady state as without feedback. Thus, as long as the coherent control loop does not have access to any 
external sources of energy, it is only able to counteract dissipation on a transient time-scale except one allows for 
perfect mirrors, which in turn would make it unnessary to introduce any feedback loop at all in our situation. It should 
be noted, however, that for transient time-scales coherent feedback might have strong advantages or it might be the case 
that one is not primarily interested in the prevention of dissipation (in fact, in Ref. \cite{GrimsmoEtAlNJP2014} they use 
the control loop to \emph{speed up} dissipation). The question whether one scheme is superior to the other is thus, in 
general, undecidable and needs a thorough case to case analysis.

\section{Feedback Scheme II}
\label{sec feedback scheme 2}

We wish to present a second feedback scheme, in which we replace the time-delayed signal by a fixed reference signal such that 
no time-delayed noise enters the description and hence, we are not forced to work with a SME like Eq. (\ref{eq SME fb scheme 1}). 
The measurement signal we wish to couple back is thus of the form 
\begin{equation}
 \delta I(t) = [I(t) - x^*(t)]dt = [\langle x\rangle(t) - x^*(t)]dt + \sqrt{\frac{\hbar}{2\gamma\eta}} dW(t)
\end{equation}
and our aim is to \emph{synchronize} the motion of the HO with the external reference signal $x^*(t)$. Choosing 
$\h z = k\h p$ and using Eq. (\ref{eq SME influence fb}) yields the SME 
\begin{equation}
 \begin{split}
  \h\rho_c(t+dt)	=&~	 \h\rho_c(t)	\\
			&+	dt \left\{\C L_0 + \C L_\text{meas} - \frac{ik}{2\hbar}\C C[\h p]\{\C A[\h x] - 2x^*(t)\} - \frac{k^2}{4\hbar\gamma\eta}\C C^2[\h p]\right\} \h\rho_c(t)	\\
			&+	dW(t) \left\{\sqrt{\frac{\gamma\eta}{2\hbar}}\C H[\h x] - \frac{ik}{\sqrt{2\hbar\gamma\eta}}\C C[\h p]\right\} \h\rho_c(t).
 \end{split}
\end{equation}
The associated FPE (\ref{eq FPE generic}) for the conditional Wigner function is given by 
\begin{equation}
 \begin{split}\label{eq W scheme 2}
  W_c(x,p,t+dt)	=&~	W_c(x,p,t) + dt \left(-\nabla^T\cdot\textbf{d} + \frac{1}{2}\nabla^T\cdot D\cdot\nabla\right) W_c(x,p,t)	\\
		&+	dW(t)\left[\sqrt{\frac{2\gamma\eta}{\hbar}}(x - \lr x_c) - k\sqrt{\frac{\hbar}{2\gamma\eta}} \frac{\partial}{\partial x}\right] W_c(x,p,t)
 \end{split}
\end{equation}
with the nonvanishing coefficients 
\begin{equation}
 \begin{split}\label{eq coeff W scheme 2}
  & \textbf{d}_x = \omega p - \frac{\kappa}{2} x + k[x-x^*(t)], ~~~ \textbf{d}_p = -\omega x - \frac{\kappa}{2} p,	\\
  & D_{xx} = \kappa\hbar\frac{1+2n_B}{2} + \frac{\hbar k^2}{2\gamma\eta}, ~~~ D_{pp} = \kappa\hbar\frac{1+2n_B}{2} + \frac{\hbar\gamma}{2}.
 \end{split}
\end{equation}
Because we have no time-delayed noise here, the average, unconditional evolution of the system can be simply obtained 
by dropping all terms proportional to the noise $dW(t)$ due to Eq. (\ref{eq Ito average}). We thus have a fully Markovian feedback 
scheme here. 

\begin{figure}
 \includegraphics[width=0.70\textwidth,clip=true]{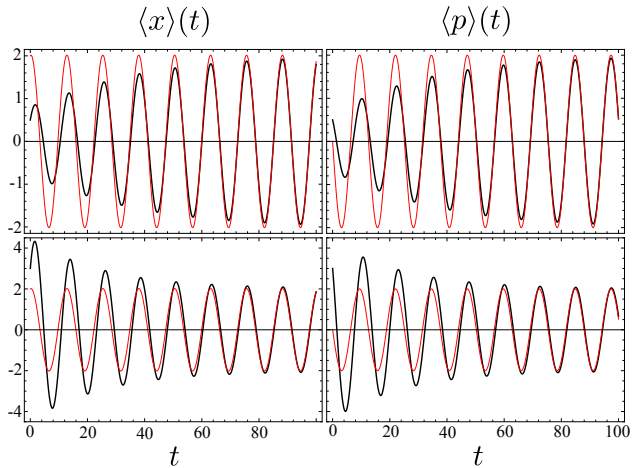}
 \label{fig plot scheme 2}
 \caption{Plot of the average mean values $\lr x(t)$ and $\lr p(t)$ as a function of time $t$ (black, thick lines) compared 
 to the asymptotic solution given by Eqs. (\ref{eq asymp sol x scheme 2}) and (\ref{eq asymp sol p scheme 2}) (red, thin lines). 
 The upper panel corresponds to the initial condition $(\lr x,\lr p)(0) = (1/2,1/2)$ and the lower one to $(\lr x,\lr p)(0) = (3,3)$. 
 Other parameters are $\omega = 1, \kappa = 1/4$ and $y_0 = 2$. }
\end{figure}

The equation of motion for the conditional means are 
\begin{align}
 d\lr x_c	=&~	\left\{\omega\lr p - \frac{\kappa}{2}\lr x + k[\lr x-x^*(t)]\right\} dt	\\
		&+	\left(\sqrt{\frac{2\gamma\eta}{\hbar}}V_{x,c} + k\sqrt{\frac{\hbar}{2\gamma\eta}} \right) dW(t),	\nonumber	\\
 d\lr p_c	=&~	\left\{-\omega\lr x - \frac{\kappa}{2}\lr p\right\} dt + \sqrt{\frac{2\gamma\eta}{\hbar}}C_c dW(t)
\end{align}
from which the average evolution directly follows: 
\begin{align}
 \frac{d}{dt}\langle x\rangle	&=	\omega\lr p - \frac{\kappa}{2}\lr x + k[\lr x-x^*(t)],	\\
 \frac{d}{dt}\langle p\rangle	&=	-\omega\lr x - \frac{\kappa}{2}\lr p.
\end{align}
Again, it is not our purpose to investigate these equations in detail, but 
we will only focus on the special situation $k = \kappa/2$ and $x^*(t) = - y_0 \cos(\omega t)$. Then, 
\begin{align}
 \frac{d}{dt}\langle x\rangle	&=	\omega\lr p + \frac{y_0\kappa}{2} \cos(\omega t),	\label{eq mean x scheme 2}	\\
 \frac{d}{dt}\langle p\rangle	&=	-\omega\lr x - \frac{\kappa}{2}\lr p.	\label{eq mean p scheme 2}
\end{align}
These equations look very similar to the classical differential equation of an externally forced harmonic 
oscillator.\footnote{Indeed, 
if we would choose the feedback operator $\h z = k\h x$, the resulting differential equations for $\langle x\rangle$ 
and $\langle p\rangle$ would exactly resemble the differential equation of a classical harmonic oscillator with 
sinusoidal driving force. } However, it is important to emphasize that we do not have an open-loop control scheme here 
although it looks like it at the average level of the means. 
The asymptotic solution of Eqs. (\ref{eq mean x scheme 2}) and (\ref{eq mean p scheme 2}) is given by 
\begin{align}
 \lim_{t\rightarrow\infty} \lr x(t)	&=	y_0\cos(\omega t) + \frac{\kappa y_0}{2\omega}\sin(\omega t),	\label{eq asymp sol x scheme 2}	\\
 \lim_{t\rightarrow\infty} \lr p(t)	&=	-y_0 \sin(\omega t).	\label{eq asymp sol p scheme 2}
\end{align}
Within the weak-coupling regime it is natural to assume that $\kappa/\omega \ll 1$ and we asymptotically obtain a circular 
motion $(\lr x,\lr p)(t) \approx y_0(\cos \omega t,-\sin\omega t)$. It is worth to stress that we always reach the asymptotic 
solution independent of the chosen initial condition, also see Fig. \ref{fig plot scheme 2}. 
As a consequence, the limit cycle given by Eqs. (\ref{eq asymp sol x scheme 2}) and (\ref{eq asymp sol p scheme 2}) is stable. 
In contrast, the equations of motion for the first scheme are completely scale-invariant, i.e., an arbitrary scaling of the form 
$(\lr{\tilde x},\lr{\tilde p}) \equiv \alpha (\lr x,\lr p), \alpha\in\mathbb{R}$, leaves the Eqs. (\ref{eq mean x scheme 1}) and 
(\ref{eq mean p scheme 1}) unchanged and the effect of the feedback depends on the initial condition. 

Within the Gaussian assumption, the conditional covariances (i.e., the covariances an observer \emph{with} access to the 
measurement result would associate to the state of the system) evolve as for the first scheme according to 
\begin{align}
 \frac{d}{dt} V_{x,c}	&=	-\kappa V_{x,c} + 2\omega C_c - \frac{2\gamma\eta}{\hbar} V_{x,c}^2 + \kappa\hbar\frac{1+2n_B}{2},	\label{eq Vxc scheme 2}	\\
 \frac{d}{dt} V_{p,c}	&=	-\kappa V_{p,c} - 2\omega C_c - \frac{2\gamma\eta}{\hbar} C_c^2 + \kappa\hbar\frac{1+2n_B}{2} + \frac{\hbar\gamma}{2},	\\
 \frac{d}{dt} C_c	&=	\omega(V_{p,c} - V_{x,c}) - \kappa C_c - \frac{2\gamma\eta}{\hbar} V_{x,c}C_c.	\label{eq Cc scheme 2}
\end{align}
In contrast, the unconditional covariances (which an observer \emph{without} access to the measurement results would 
associate to the state of the system) obey 
\begin{align}
 \frac{d}{dt}V_x	&=	(2k-\kappa)V_x + 2\omega C + \kappa\hbar\frac{1+2n_B}{2} + \frac{\hbar k^2}{2\gamma\eta},	\label{eq Vx scheme 2}	\\
 \frac{d}{dt}V_p	&=	-\kappa V_p - 2\omega C + \kappa\hbar\frac{1+2n_B}{2} + \frac{\hbar\gamma}{2},	\\
 \frac{d}{dt}C		&=	\omega(V_p-V_x) + (k-\kappa)C.	\label{eq C scheme 2}
\end{align}
Comparing both sets of equations, the most striking difference is that Eqs. (\ref{eq Vxc scheme 2}) -- (\ref{eq Cc scheme 2}) 
are nonlinear differential equations whereas Eqs. (\ref{eq Vx scheme 2}) -- (\ref{eq C scheme 2}) are linear. Especially 
note the term proportional to $- \frac{2\gamma\eta}{\hbar} V_{x,c}^2$ in Eq. (\ref{eq Vxc scheme 2}), which tends to squeeze 
the wavepacket in the $x$-direction. This is the effect of the continuous measurement performed on the system, which tends to localize 
the state. However, if we average over (or, equivalently, ignore) the measurement results, this effect is missing. 
Furthermore, note that Eqs. (\ref{eq Vxc scheme 2}) -- (\ref{eq Cc scheme 2}) do not contain the parameter $k$, 
which quantifies how strongly we feed back the signal. 

Solving Eqs. (\ref{eq Vxc scheme 2}) -- (\ref{eq Cc scheme 2}) for its steady state is possible, but the exact expressions are 
extremely lengthy. However, to see how the continuous measurement influences the conditional covariances we will have a look at 
the special case of no damping ($\kappa = 0$). To appreciate this case we remind us that the steady state covariances of a damped 
HO for no measurement and no feedback are given by $V_x = V_p = \hbar(n_B + \frac{1}{2})$ and $C = 0$.\footnote{We can obtain this 
result by computing the steady state of Eqs. (\ref{eq Vx scheme 2}) -- (\ref{eq C scheme 2}) where we first send $k\rightarrow0$ and 
then $\gamma\rightarrow0$. } Especially, at zero temperature ($n_B = 0$), we have a minimum uncertainty 
wave packet satisfying the lower bound of the Heisenberg uncertainty relation, $V_xV_p = \hbar^2/4$. 
Now, for $\kappa = 0$, we can expand the conditional covariances in powers of $\gamma$: 
\begin{align}
 \lim_{t\rightarrow\infty} V_{x,c}(t)	&=	\frac{\hbar}{2\sqrt{\eta}} - \frac{\sqrt{\eta}\hbar\gamma^2}{16\omega^2} + \C O(\gamma^3),	\\
 \lim_{t\rightarrow\infty} V_{p,c}(t)	&=	\frac{\hbar}{2\sqrt{\eta}} + \frac{3\sqrt{\eta}\hbar\gamma^2}{16\omega^2} + \C O(\gamma^3),	\\
 \lim_{t\rightarrow\infty} C_{c}(t)	&=	\frac{\hbar\gamma}{4\omega} + \C O(\gamma^3).
\end{align}
We see that the uncertainty in position is reduced at the expense of an increased uncertainty in momentum. This is exactly what 
we have to expect from a position measurement due to Heisenberg's uncertainty principle. Note that this effect is weakened for larger 
frequencies $\omega$ of the oscillator, because the continuous measurement has problems to ``follow'' the state appropriately. 
Furthermore, an imperfect detector ($\eta<1$) will always increase the variances. Finally, we remark that these results become 
meaningless in the strict limit $\gamma\rightarrow0$, because in that case there is simply \emph{no} conditional dynamics. This 
becomes also clear by looking at Eq. (\ref{eq Vx scheme 2}), in which the last term diverges in this limit, because we would feed 
back an infinitely noisy signal. 

Thus, an observer with access to the measurement record would associate very different covariances to the system in comparison 
to an observer without that knowledge. However, a detailed discussion of the time evolution of the covariances is beyond the scope 
of the present paper. Instead, we find it more interesting to discuss the relationship between the present quantum feedback scheme and 
its classical counterpart, to which we turn now.

\section{Classical limit}
\label{sec classical limit}
\index{classical limit}\index{Fokker-Planck equation, stochastic}\index{Fokker-Planck equation}

To take the classical limit in our case we have to use a little trick, because simply taking $\hbar\rightarrow0$ does not 
yield the correct result. In fact, for $\hbar\rightarrow0$ we have from Eq. (\ref{eq meas record}) that 
$I(t) = \langle x\rangle_c(t)$, which only makes sense if we can observe the particle with infinite accuracy, i.e., its 
conditional probability distribution is a delta function with respect to the position $x$. 
We explicitly wish, however, to model a \emph{noisy} classical measurement. We thus additionally demand that $\eta\rightarrow0$. 
More specifically, we set $\eta \equiv \hbar/\sigma$ with $\sigma$ finite such that Eq. (\ref{eq meas record}) becomes 
\begin{equation}\label{eq meas record classical}
 dI(t) = \langle x\rangle_c(t) dt + \sqrt{\frac{\sigma}{2\gamma}} dW(t)
\end{equation}
and we remark that the case $\sigma\rightarrow0$ corresponds to an error-free measurement. 

The FPE (\ref{eq FPE generic}) for the free evolution of the HO with drift vector and diffusion matrix from Eq. 
(\ref{eq W drift diffusion free HO}) becomes for $\hbar\rightarrow0$ (note that the Bose-Einstein distribution $n_B$ 
contains $\hbar$ as well and needs to be expanded) 
\begin{equation}\label{eq FPE free}
 \begin{split}
  \frac{\partial}{\partial t}P(x,p,t)	&\equiv	\C L_0^\text{cl} P(x,p,t)	\\
					&=	\left\{-\nabla^T\cdot \binom{\omega p - \frac{\kappa}{2} x}{-\omega x - \frac{\kappa}{2} x} + \frac{\kappa}{2\beta\omega}\nabla^T\cdot\nabla\right\}P(x,p,t).
 \end{split}
\end{equation}
To emphasize the fact that the Wigner function $W(x,p)$ becomes an ordinary probability distribution in the classical limit, 
we denoted it by $P(x,p)$. As expected, we see that Eq. (\ref{eq FPE free}) corresponds to a FPE for 
a Brownian particle in a harmonic potential where position and momentum are both damped (usually one considers only the 
momentum to be damped \cite{RiskenBook1984}). This peculiarity is a consequence of an approximation made in deriving the ME 
(\ref{eq ME damped HO}), which is known as the secular or rotating-wave approximation. Nevertheless, one easily confirms 
that the canonical equilibrium state $P_\text{eq} \sim \exp[-\beta\omega(p^2+x^2)/2]$ is a steady state of this FPE as it must be. 

Next, it turns out to be interesting to discuss the classical limit of the SME (\ref{eq SME free plus meas}) describing the 
free evolution plus the influence of the continuous noisy measurement. Using Eq. (\ref{eq W meas}) we obtain an equation 
for the conditional probability distribution $P_c(x,p)$ 
\begin{equation}\label{eq FPE free plus meas}
 P_c(x,p,t+dt) = \left\{1 + \C L_0^\text{cl} dt + \sqrt{\frac{2\gamma}{\sigma}}(x-\langle x\rangle_c) dW(t)\right\}P_c(x,p,t).
\end{equation}
This is a \emph{stochastic} FPE, which is nonlinear in $P_c$. It describes how our state of knowledge changes 
if we take into account the measurement record (\ref{eq meas record classical}). However, averaging Eq. 
(\ref{eq FPE free plus meas}) over all measurement results yields Eq. (\ref{eq FPE free}), which reflects the fact that 
a classical measurement does not perturb the system.\footnote{This is 
true at least in our context. In principle, it is of course possible to construct classical measurements, which perturb 
the system, too \cite{WisemanMilburnBook2010}. } 
This is in contrast to the quantum case where the average evolution is still influenced by $\C L_\text{meas}$, 
see Eq. (\ref{eq ME av meas}). Hence, the term $\C L_\text{meas}$ in Eq. (\ref{eq ME av meas}) is \emph{purely} of quantum 
orgin and it describes the effect of decoherence on a quantum state under the influence of a measurement. This effect 
is absent in a classical world. Exactly the same equation 
and the same conclusions were already derived by Milburn following a different route \cite{MilburnQSO1996}. 

The impact of these conclusions is, however, much more severe if one additionally considers feedback. As we will now show, 
applying feedback based on the use of the stochastic FPE (\ref{eq FPE free plus meas}), \emph{does} indeed yield observable 
consequences even \emph{on average}. Please note that trying to model the present situation by a classical Langevin 
equation is nonsensical. If we would use a Langevin equation to describe our state of knowlegde about the system, 
we would implicitly ascribe an objective reality to the fact that there \emph{is} a definite position $x_0$ 
and momentum $p_0$ of the particle corresponding to a probability distribution $\delta(x-x_0)\delta(p-p_0)$. 
This is, however, \emph{not true} from the point of view of the observer who has to apply feedback control based on 
incomplete information (i.e., the noisy measurement record). Results we would obtain from a Langevin equation treatment 
can only be recovered in the limit of an error-free measurement, i.e., for $\sigma\rightarrow0$, as we will demonstrate 
in appendix \ref{sec appendix}. 

For simplicity we will only have a look at the second feedback scheme from Sec. \ref{sec feedback scheme 2}, because we 
can directly obtain the classical limit for the average evolution from Eq. (\ref{eq W scheme 2}). The same situation is, 
however, also encountered by considering the first scheme. Taking $\hbar\rightarrow0$ in Eq. (\ref{eq W scheme 2}) 
together with the coefficients (\ref{eq coeff W scheme 2}) then yields the FPE 
\begin{equation}
 \frac{\partial}{\partial t}P(x,p,t) = \left\{\C L_0^\text{cl} - \partial_x k[x-x^*(t)] + \frac{k^2\sigma}{2\gamma} \partial_x^2\right\} P(x,p,t).
\end{equation}
The first term to the correction of $\C L_0^\text{cl}$ is the term one would expect for a noiseless feedback loop, too. 
The second, however, only arises due to the noisy measurement (we see that it vanishes for $\sigma\rightarrow0$) 
and causes an additional diffusion in the $x$ direction simply due to the fact that the observer applies 
a slightly wrong feedback control compared to the ``perfect'' situation without measurement errors. 

We thus conclude this section by noting that the treatment of continuous noisy classical measurements faces similar 
challenges as in the quantum setting. On average the measurement itself does not influence the classical dynamics, but we 
see that we obtain new terms even on average if we use this measurement to perform feedback control. Most importantly, 
because a feedback loop has to be implemented by the observer who has access to the measurement record, it is in general 
not possible to model this situation with a Langevin equation. Furthermore, we remark that the situation is expected to be 
even more complicated for time-delayed feedback, where no average description is a priori possible.

\section{Summary and outlook}
\label{sec discussion}

Because we discussed the meaning of our results already during the main text in detail, we will only give a short summary 
together with a discussion on possible extensions and applications. 

We have used two simple feedback schemes, which are known to change the stability of a steady state obtained from 
linearizing a dynamical system around that fixed point in the classical case. 
For the simple situation of a damped quantum HO we have seen that \emph{on average} we obtain the same dynamics 
for the mean values as expected from a classical treatment and thus, classical control strategies might turn out to be 
very useful in the quantum realm, too. 

However, the fact that a classical control scheme works so well in the quantum regime depends on two crucial assumptions. 
First of all, we have used a linear system (the HO). Having a non-linear system Hamiltonian (e.g., a Hamiltonian with 
a quartic potential $\sim \h x^4$) would complicate the treatment, because already the equations for the mean values 
would contain higher order moments, as e.g. $\lr{x^3}$ in case of the quartic oscillator. Simply factorizing them 
as $\lr{x^3} \approx \lr x^3$ would imply that we are already using a classical approximation. However, 
as in the classical treatment, where the equations of motion are obtained from linearizing a (potentially non-linear) 
dynamical system around the fixed point, it might also be possible in the quantum regime to neglect non-linear terms 
in the vicinity of the fixed point. Whether or not this is possible crucially depends on the localization 
of the state in phase space, i.e., on its covariances. Here, continuous quantum measurements can actually turn out 
to be helpful, because they tend to localize the wavefunction and counteract a possible spreading of the state. 

The second important assumption we used was that we restricted ourselves to continuous variable quantum systems. 
The reason why we obtained simple equations of motion is related 
to the commutation relation $[\h x,\h p] = i\hbar$, which we implicitly used to obtain the evolution equation for 
the Wigner function. Formally, phase space methods are also possible for other quantum systems, but the maps are much 
more complicated \cite{CarmichaelBook1999a}. For such systems the methods presented here might be useful under certain special 
assumptions, but in general one should expect them to fail.

\section*{Acknowledgments}

PS wishes to thank Philipp H\"ovel, Lina Jaurigue and Wassilij Kopylov for helpful discussions about time-delayed feedback 
control. Financial support by the DFG (SCHA 1646/3-1, SFB 910, and GRK 1558) is gratefully acknowledged.


\bibliographystyle{spphys}
\bibliography{refs_SFB_book_2015}


\appendix

\section{Appendix}
\label{sec appendix}
\index{Langevin equation}\index{Fokker-Planck equation, stochastic}\index{Fokker-Planck equation}

We want to show that the stochastic FPE, which in general describes the \emph{incomplete} state of knowledge 
of an observer, reduces to a Langevin equation in the error-free limit, i.e., in the limit in which we have indeed 
\emph{complete} knowledge about the state of the system. Because we are only interested in a proof of principle here, 
we will consider the simplified situation of an overdamped particle.\footnote{The complete description of an 
underdamped particle (i.e., a particle descibed by its position $x$ \emph{and} momentum $p$), which is based on a 
continuous measurement of its position $x$ alone, Eq. (\ref{eq meas record classical}), faces the additional 
challenge that we have to first estimate the momentum $p$ based on the noisy measurement results. } 
The Langevin equation of an overdamped Brownian particle in an external potential $U(x)$ is usually given as 
(see, e.g., \cite{RiskenBook1984}) 
\begin{equation}\label{eq Langevin}
 \dot x(t) = -\frac{2}{\kappa}U'(x) + \sqrt{\frac{T}{\kappa}} \xi(t)
\end{equation}
with $U'(x) \equiv \frac{\partial U(x)}{\partial x}$ and the Gaussian white noise $\xi(t) \equiv \frac{dW(t)}{dt}$. 
Furthermore, note that our friction constant is $\frac{\kappa}{2}$ and not -- as it is often denoted -- $\gamma$, 
because we use $\gamma$ already for the measurement rate. Now, 
it is important to remark that at this point Eq. (\ref{eq Langevin}) simply describes a convenient 
\emph{numerical tool to simulate} a stochastic process. By the mathematical rules of stochastic calculus it is 
guaranteed that the Langevin equation gives the same averages as the corresponding FPE, i.e., the ensemble average 
$\mathbb{E}[f(x)]$ over all noisy trajectories for some function $f(x)$ is equal to the expectation value 
$\langle f(x)\rangle$ taken with respect to the solution of the FPE. 

In Sec. \ref{sec classical limit} we have suggested that the correct state of the system based on a noisy position 
measurement is given by the stochastic FPE (\ref{eq FPE free plus meas}), which for an overdamped particle becomes 
(see also Ref. \cite{MilburnQSO1996}) 
\begin{equation}\label{eq FPE appendix}
 P_c(x,t+dt) = \left\{1 + \C L_0^\text{cl} dt + \sqrt{\frac{2\gamma}{\sigma}}(x-\langle x\rangle_c) dW(t)\right\}P_c(x,t)
\end{equation}
with \cite{RiskenBook1984} 
\begin{equation}
 \C L_0^\text{cl} = \frac{\partial}{\partial x}\left(\frac{2U'(x)}{\kappa} + \frac{\partial}{\partial x}\frac{2T}{\kappa}\right).
\end{equation}
Furthermore, we have also claimed that the parameter $\sigma$ in (\ref{eq meas record classical}) quantifies the error 
of the measurement. This suggests that we should be able to recover the Langevin Eq. (\ref{eq Langevin}) in the limit 
$\sigma\rightarrow0$ in which we can observe the particle with infinite precision. 

To show this we compute the expectation value of the position according to Eq. (\ref{eq FPE appendix}): 
\begin{equation}\label{eq app mean x EOM}
 d\lr x_c(t) = -\frac{2}{\kappa}\lr{U'(x)}_c dt + \sqrt{\frac{2\gamma}{\sigma}} V_c(t) dW(t)
\end{equation}
where $V_c = \lr{x^2}_c - \lr x_c^2$ denotes the variance of the particles position. Because the conditional variance 
enters this equation, we compute its time evolution, too: 
\begin{equation}
 \begin{split}\label{eq app variance EOM}
  dV_c	=&	-\frac{4}{\kappa}\left[\lr{xU'(x)}_c(t) - \lr x_c(t)\lr{U'(x)}_c(t)\right] dt + \frac{4T}{\kappa}dt - \frac{2\gamma}{\sigma}V_c^2 dt	\\
	&+	\sqrt{\frac{2\gamma}{\sigma}}\left\langle[x-\lr x_c(t)]^3\right\rangle dW(t).
 \end{split}
\end{equation}
To make analytical progress we now need two assumptions. First, we will assume that $P_c(x,t)$ is a Gaussian 
probability distribution. In fact, because the measurement tends to localize the probability distribution and it is 
itself modeled as a Gaussian process, this assumption seems reasonable. In addition, we expect $P_c(x,t)$ to become 
a delta-distribution in the limit $\sigma\rightarrow0$, which is a Gaussian distribution, too. This assumption allows 
us to drop the stochastic term in Eq. (\ref{eq app variance EOM}). Second, within the variance of $P_c(x,t)$ we 
assume that we can expand $U(x)$ in a Taylor series and approximate it by a quadratic function $ax^2 + bx +c$. 
This implies $\lr{xU'(x)}_c(t) \approx a\lr{x^2}_c(t) + b\lr x_c(t)$. In fact, this assumption seems also reasonable, 
because we expect the measurement to be precise enough such that we can locally resolve the evolution of the particle 
sufficiently well (especially for small $\sigma$); or to put it differently: a measurement only makes sense if the 
conditional variance $V_c$ of $P_c(x,t)$ is small enough. Using this approximation, too, we can then write Eq. 
(\ref{eq app variance EOM}) as 
\begin{equation}
 \frac{d}{dt}V_c(t) = \frac{4\kappa}{2} - \frac{4a}{\kappa}V_c - \frac{2\gamma}{\sigma}V_c^2.
\end{equation}
The only physical steady state solution of this equation is 
\begin{equation}
 \lim_{t\rightarrow\infty} V_c(t) = \frac{a\sigma}{\gamma\kappa}\left(\sqrt{1 + \frac{2T\gamma\kappa}{a^2\sigma}} - 1\right).
\end{equation}
Inserting this into Eq. (\ref{eq app mean x EOM}) yields 
\begin{equation}
 d\lr x_c(t) = -\frac{2}{\kappa}\lr{U'(x)}_c dt + \frac{a}{\kappa} \sqrt{\frac{2\sigma}{\gamma}} \left(\sqrt{1 + \frac{2T\gamma\kappa}{a^2\sigma}} - 1\right) dW(t).
\end{equation}
In this equation we can take the limit $\sigma\rightarrow0$ such that 
\begin{equation}
 d\lr x^0_c(t) = -\frac{2}{\kappa}\lr{U'(x)}^0_c dt + \sqrt{\frac{4T}{\kappa}} dW(t),
\end{equation}
where we introduced a superscript $0$ on all expectation values to denote the error-free limit. This equation looks 
already very similar to the LE (\ref{eq Langevin}). In fact, mathematically \emph{this is the LE} since from 
Eq. (\ref{eq meas record classical}) we can see that the measurement result becomes for $\sigma\rightarrow0$ 
$dI(t) = \lr x_c^{0}(t) dt$. This implies that the measurement result is deterministic and not stochastic 
anymore, which is only compatible if the associated probability distribution $P_c(x,t)$ is a delta distribution 
$\delta(x-x^*)$ where $x^*$ describes the \emph{true instantaneous position} of the particle without any uncertainty. 
Then, Eq. (\ref{eq app mean x EOM}) becomes 
\begin{equation}
 d x^*(t) = -\frac{2}{\kappa}U'(x^*) dt + \sqrt{\frac{4T}{\kappa}} dW(t).
\end{equation}
Now, this equation is \emph{not just a numerical tool, but describes real physical objectivity}, because $x^*$ coincides 
with the observed position in the lab. This distinction might seem very nitpicking, but it is of crucial importance if 
we want to perform feedback based on incomplete information.

\end{document}